\documentclass[a4paper]{jpconf}
\usepackage{graphicx}

\usepackage{amsmath, amssymb}
\usepackage{color}
\usepackage{bm}
\usepackage{mathrsfs}

\newcommand{\diff}{\mathrm{d}}

\newcommand{\imag}{\mathrm{Im}\,}
\newcommand{\imu}{\mathrm{i}}
\newcommand{\epn}{\mathrm{e}}

\newcommand{\ua}{\uparrow}
\newcommand{\da}{\downarrow}
\newcommand{\dg}{\dagger}
\newcommand{\la}{\langle}
\newcommand{\ra}{\rangle}
\newcommand{\al}{\alpha}
\newcommand{\sg}{\sigma}

\newcommand{\ep}{\varepsilon}

\begin{document}
\title{Collective excitations from 
composite orders
in Kondo lattice with non-Kramers doublets
}

\author{S Hoshino$^1$ and Y Kuramoto$^2$}

\address{$^1$ Department of Basic Science, The University of Tokyo, Meguro 153-8902, Japan}
\address{$^2$ Department of Physics, Tohoku University, Sendai 980-8758, Japan}

\ead{kuramoto@cmpt.phys.tohoku.ac.jp.ac.jp}

\begin{abstract}
Goldstone modes emerge associated with spontaneous breakdown of the continuous symmetry in the two-channel Kondo lattice, which describes strongly correlated $f$-electron systems with a non-Kramers doublet at each site.
This paper derives the spectra of these collective modes by the equation of motion method together with the random phase approximation.
The diagonal composite order breaks the SU(2) channel symmetry, and the symmetry-restoring collective mode couples with magnetic field.
On the other hand, 
the off-diagonal or superconducting composite 
order breaks the gauge symmetry of conduction electrons, and the collective mode couples with charge excitations near the zone boundary. 
At half-filling of the conduction bands, the spectra of these two modes become identical by a shift of the momentum,
owing to the SO(5) symmetry of the system.
The velocity of each Goldstone mode involves not only the Fermi velocity of conduction electrons but amplitude of 
the mean field
as a multiplying factor.
Detection of the Goldstone mode should provide 
a way to identify the composite order parameter.
\end{abstract}

\section{Introduction}
Nearly localized $f$ electrons in a metal interact with conduction electrons to cause a variety of interesting behaviors.
By hybridization between $f$ and conduction electrons,
each $f$ electron tends to form a spin-singlet state with conduction electrons, which is called the Kondo effect.
If the crystalline-electric-field ground state 
makes a doublet with even number of $f$ electrons per site, as in 
Pr$^{3+}$- or U$^{4+}$-based systems, the two degenerate states 
is not connected by the time reversal, in contrast with a single spin.
Such pair of states are called a non-Kramers doublet.
It is known that the Kondo effect 
in the impurity system with
a non-Kramers doublet causes
peculiar phenomena such as a non-Fermi liquid ground state \cite{nozieres, cox, cox98},
where the entropy remains finite even at zero temperature.
In contrast with a single impurity, the lattice of these non-Kramers doublets
should undergo some ordering to remove the entropy of the whole system.
There has been a long-standing interest in the resultant electronic order, especially in relation to exotic behaviors
in non-Kramers systems such as UBe$_{13}$ \cite{ott, shimizu}, and more recently in 
PrIr$_{2}$Zn$_{20}$ and PrTi$_{2}$Al$_{20}$ \cite{onimaru, sakai, matsubayashi, tsujimoto}.
The common features observed in these systems are
the non-Fermi liquid behavior and superconductivity at low temperatures.

To investigate the characteristic electronic order
in non-Kramers doublet systems, we take the two-channel Kondo lattice as the simplest model \cite{emery93, jarrell97}.
We have already demonstrated \cite{hoshino11,hoshino13,hoshino14}
that the two-channel Kondo effect realizes exotic 
ground states with ``composite order'', where the order parameter is not described by one-body 
quantities
 such as magnetization or density, but by combination of $f$- and conduction-electron degrees of freedom.
The diagonal composite order is identified as itinerant multipoles, and the off-diagonal one is a superconducting state with staggered order parameter.
It has also been shown that
a composite order is equivalent to ``odd-frequency order'' with a one-body but time-dependent order parameter of conduction electrons \cite{emery92, balatsky}.
This concept is useful especially in detection of the instability toward composite orders.
Furthermore, one of the authors \cite{hoshino14-2} has recently 
given a simple one-body picture of the orders by using a mean-field theory together with consideration of the SO(5) symmetry.

In this paper we focus on collective excitation (Goldstone) modes that emerge from both diagonally and off-diagonally ordered states.  These collective modes are gapless because the broken symmetry in either case is continuous;  SU(2) in the diagonal order, and U(1) in the off-diagonal order. 
The detection of the peculiar Goldstone mode should provide clear evidence of the composite order.
The collective modes couples either with magnetic field (diagonal order), or electric field (off-diagonal order).  
Hence powerful experimental probes such as neutrons, ultrasound or light can be used for the detection.

In deriving the spectrum of the collective modes, we employ a mean-field theory and the random phase approximation. 
Before discussing the spectrum, we begin in the next section
with reviewing the electronic orders and their symmetry in the 
two-channel Kondo lattice. 
In Sec. 3, we approach the ordered states from normal state at high temperatures by looking at the relevant response function.
Although the response function probing the composite order is extremely complicated, 
the transition temperature can be derived by exploiting the equivalence to the odd-frequency one-body order.
In Sec. 4 we proceed to the mean-field description of odd-frequency orders.
The diagonally ordered state has a broken channel symmetry, 
and the corresponding mean-field is a hybridization between fictitious local fermions and conduction electrons in a particular channel.
This hybridization results in a Kondo insulating state for one of the two channels, while the other channel remains a Fermi liquid.
In the superconducting state, on the other hand, it is essential that we have two kinds of mean fields; hybridization in a particular channel, and pairing with fictitious $f$ electrons in the other channel.
With these preliminaries, we derive in Sec. 5 the spectra of 
the Goldstone modes both in diagonally and off-diagonally ordered phases. 
The final section summarizes the paper, and discusses relevance of our results to actual materials.

\section{Composite order parameters in Kondo lattice with non-Kramers doublets}

Let us consider a non-Kramers $\Gamma_3$ doublet 
in cubic symmetry for the ground state of the $f^2$ configuration.
We regard the doublet as pseudospin states described by the operator $\bm S$.
We assume that they couple to conduction electrons with the $\Gamma_8$ representation, which have both pseudospin ($\sg$) and real spin ($\al$) degrees of freedom.
This system is described by the two-channel Kondo lattice Hamiltonian given by
\begin{align}
{\cal H} = - t \sum_{\la ij \ra\al\sg} c^\dg_{i\al\sg } c_{j\al\sg} 
- \mu \sum_{i\al\sg} c^\dg_{i\al\sg } c_{i\al\sg}
+ J \sum_{i\al} \bm S_i \cdot \bm s_{ci\al}
,
\label{2chKL}
\end{align}
where $c_{i\al\sg}$ is the annihilation operator of conduction electron with channel $\al=1,2$ and pseudospin $\sg=\ua,\da$ at site $i$.
The summation with respect to $\la ij \ra$ is taken over the nearest-neighbor sites on the bipartite lattice.
We have also introduced pseudospin operator for conduction electrons as $\bm s_{ci\al} = \tfrac 12 \sum_{\sg\sg'} c^\dg_{i\al\sg} \bm \sg_{\sg\sg'} c_{i\al\sg'}$.
Note that the channel $\al$ physically describes the real spin (or magnetic) degrees of freedom.

The disordered state of this model cannnot be a Fermi liquid, since the localized spins are overscreened by conduction electrons due to the presence of two channels \cite{cox98}.
This means that the disordered state has a residual entropy and cannot be the ground state.
The simplest way to resolve the entropy is an ordering by the RKKY interaction among the localized pseudospins.  
In our previous studies, we have also found 
other peculiar ordered states with composite order parameters.
We have both diagonal and off-diagonal composite orders, each of which is characterized by the following two-body quantities \cite{hoshino14}:
\begin{align}
\Psi^z &= \sum_{i\al\al'\sg\sg'} c^\dg_{i\al\sg}
\sigma^z_{\al\al'}[\bm S_i \cdot \bm \sg_{\sg\sg'}]
c_{i\al'\sg'}
=
2 \sum_{i} \bm S_{i} \cdot (\bm s_{ci1} - \bm s_{ci2})
,\label{Psi^z}
\\
\Phi^+ &= 
\sum_{i\al\al'\sg\sg'} c^\dg_{i\al\sg}
\epsilon_{\al\al'}[\bm S_i \cdot (\bm \sg \epsilon)_{\sg\sg'}]
c^\dg_{i\al'\sg'}
\, \epn^{\imu \bm Q\cdot \bm R_i}
\end{align}
where $\epsilon = \imu \sg^y$ is the antisymmetric unit tensor.
Here $\Psi^z$ describes
the channel symmetry breaking, and equivalent quantities
$\Psi^x$ and $\Psi^y$ can be defined by changing the component of the Pauli matrix.
On the other hand, $\Phi^+$ describes
the channel-singlet/pseudospin-singlet pairing, and the equivalent quantity
$\Phi^- \equiv (\Phi^+)^\dagger$ can be introduced.
The phase factor $\exp(\imu \bm Q\cdot \bm R_i)$ 
becomes $\pm 1$ according to whether $\bm R_i$ belongs to A or B sublattice, and 
describes the staggered order.
In the three-dimensional cubic lattice, for example, we take $\bm Q = (\pi, \pi, \pi)$ with lattice constant set to unity.

These diagonal and off-diagonal orderings with five (3+2) independent components are degenerate at half filling.
To demonstrate the degeneracy, let us 
introduce a unitary transformation $\mathscr{U}$ as follows:
\begin{align}
\mathscr{U} c_{i1\sg} \mathscr{U}^{-1} &= \frac 1 {\sqrt 2}
\sum_{\sg'} (\delta_{\sg\sg'}c_{i1\sg} + \epsilon_{\sg\sg'} \epn^{\imu \bm Q \cdot \bm R_i} c^\dg_{i2\sg'} ) 
, \label{eq:unitary_transf1}\\
\mathscr{U} c_{i2\sg} \mathscr{U}^{-1} &= \frac 1 {\sqrt 2}
\sum_{\sg'} (\delta_{\sg\sg'}c_{i1\sg} - \epsilon_{\sg\sg'} \epn^{\imu \bm Q \cdot \bm R_i} c^\dg_{i2\sg'} ) 
, \label{eq:unitary_transf2}\\
\mathscr{U} \bm S_{i} \mathscr{U}^{-1} &= \bm S_{i}
, \label{eq:unitary_transf3}
\end{align}
which leaves the Hamiltonian invariant at half filling: $[\mathscr{U}, {\cal H}_{\mu=0}] = 0$.
On the other hand, we obtain the relation
\begin{align}
\mathscr{U} \Psi^z \mathscr{U}^{-1} &= \Phi^x
,
\end{align}
where $\Phi^x = (\Phi^+ + \Phi^-)/2$.
Thus it is shown that the ordered states with $\Psi^z $ and $\Phi^x$ are degenerate.
In a similar manner, 
we can demonstrate the degeneracy among other components by
a rotation 
in the five-dimensional space of the order parameters, which is identical to the SO(5) representation or its isomorphic Sp(2) representation \cite{affleck}.

For general filling with $\mu\neq 0$, we no longer have the particle-hole symmetry, and the diagonal and off-diagonal orders have different transition temperatures. 
It has been demonstrated numerically that the superconducting order is the most stable for a density range off half-filling \cite{hoshino14}.
In the rest of the paper, we concentrate on the half-filled case with $\mu=0$, neglecting the trivial pseudospin order caused by the RKKY interaction.  
Owing to the SO(5) symmetry at $\mu=0$, information about the diagonal order gives equivalent knowledge about the off-diagonal order.

\section{Odd-frequency susceptibility and instability toward composite order}

In order to discuss the composite channel order, it is convenient to begin with the standard instability theory for the second-order phase transition in electron systems.
We first introduce a homogeneous channel moment operator by
\begin{align}
m^z(\tau, \tau') = \sum_{i\al\al'\sg} c_{i\al\sg}^\dg (\tau) \sg^z_{\al\al'}  c_{i\al'\sg} (\tau') 
,
\end{align}
where ${\cal O}(\tau) = \epn^{\tau {\cal H}} {\cal O} \epn^{-\tau {\cal H}}$ is the Heisenberg picture with imaginary time.
We allow for $\tau\neq \tau'$ for later convenience.
For the ordinary channel-ordered state, the order parameter is simply given by $\la m^z (0,0)\ra$.
Let us assume that the channel order occurs by a second-order phase transition, which is signaled by divergence of the static susceptibility:
\begin{align}
\chi^{zz} = \int_0^\beta \la
T_\tau m^z (\tau, \tau) m^z (0,0)
\ra \, \diff \tau
, \label{eq:chan_suscep}
\end{align}
where $T_\tau$ is the imaginary-time ordering operator.
The susceptibility can be rewritten in terms of the two-particle Green function.
Namely we define
\begin{align}
\chi^{zz} (\tau_1, \tau_2, \tau_3, \tau_4) &= \la
T_\tau m^z (\tau_1, \tau_2) m^z (\tau_3, \tau_4)
\ra
, \label{eq:tpgf}
\end{align}
and its Fourier transform  by
\begin{align}
\chi^{zz} (\imu\ep_n, \imu\ep_{n'}) = \frac{1}{\beta^2} \int_0^\beta \diff\tau_1\cdots \diff\tau_4 \ 
\chi^{zz} (\tau_1, \tau_2, \tau_3, \tau_4) \, 
\epn^{\imu\ep_n(\tau_2 - \tau_1)} \epn^{\imu\ep_{n'}(\tau_4 - \tau_3)}
,
\end{align}
where $\ep_n = (2n+1)\pi/\beta$ is the fermionic Matsubara frequency.
In terms of  of this quantity, Eq.~\eqref{eq:chan_suscep} can be written as
\begin{align}
\chi^{zz} = \frac{1}{\beta} \sum_{nn'} \chi^{zz} (\imu\ep_n, \imu\ep_{n'})
.
\end{align}

Let us now turn to the composite ordering where the order-parameter is a two-body quantity.  This means that the corresponding susceptibility is made of four-body quantity, which is extremely cumbersome to calculate, if not impossible.
Instead, we notice the relation
\begin{align}
\left. \frac{\partial m^z(\tau, \tau')}{\partial \tau'}
\right|_{\tau,\tau'=0}
\equiv
\tilde m^z (\tau, \tau') 
= \frac{J}{2}  \Psi^z  - t \sum_{\la ij\ra\al\al'\sg} c_{i\al\sg}^\dg \sg^z_{\al\al'} c_{j\al'\sg}
, \label{eq:deriv_chan_mom}
\end{align}
where the right-hand side (RHS) includes $\Psi^z$.
Hence the left-hand side of Eq.~\eqref{eq:deriv_chan_mom} describes the composite order parameter.
Namely, the instability toward the composite channel order can be detected from the following susceptibility:
\begin{align}
&\tilde \chi^{zz} = \int_0^\beta \la
T_\tau \tilde m^z (\tau, \tau) \tilde m^z (0,0)
\ra \, \diff \tau
= - \frac{1}{\beta} \sum_{nn'} g(\ep_n) g(\ep_{n'}) \chi^{zz} (\imu\ep_n, \imu\ep_{n'})
+ 2\la {\cal H} \ra
, \label{eq:tilde_suscep}
\end{align}
where $g(\ep_n) = \ep_n \epn^{\imu \ep_n \eta}$ with $\eta = +0$.
The form factor $g(\ep_n)$ has appeared because of the time-derivative, and the 
factor $\epn^{\imu \ep_n \eta}$ is necessary to attain the convergence.
The term $\la {\cal H} \ra$ in the RHS originates from the time-ordering operator in the two-particle Green function \eqref{eq:tpgf}, which can be neglected
as long as we are concerned about the divergence of $\tilde \chi^{zz}$.

For practical calculation, the form factor $g(\ep_n)$ as it stands is awkward since the high-frequency part in the summation has a large contribution, even though 
it remains finite at any temperature due to the presence of $\epn^{\imu \ep_n \eta}$.
Since we are concerned only with divergence of the 
Eq.~\eqref{eq:tilde_suscep}, we replace $g(\ep_n)$ by
\begin{align}
g(\ep_n) &\longrightarrow {\rm sgn\,} \ep_n
,
\end{align}
with which we no longer need the convergence factor.
In the imaginary-time domain, it is equivalent to the following replacement:
\begin{align}
\tilde \chi^{zz} \longrightarrow  -  \frac{1}{\beta^2} \int_0^\beta \diff \tau_1 \cdots \tau_4 \,
s(\tau_1 - \tau_2) s(\tau_3-\tau_4) \la T_\tau m^z (\tau_1, \tau_2) m^z (\tau_3, \tau_4) \ra
,
\end{align}
where
\begin{align}
s(\tau) 
= \imu \, {\rm cosec} ( {\pi \tau}/\beta).
\end{align}
With this replacement, the response function becomes quantitatively different from the original susceptibility.
Nevertheless, the divergence occurs at the same transition temperature because the replacement causes only a finite difference.
Since the form factor relevant to the composite order is an odd function in frequency, 
$\tilde \chi^{zz}$ is called an odd-frequency susceptibility.
To make contrast, ordinary susceptibility such as Eq.~\eqref{eq:chan_suscep} may be called even-frequency susceptibility.
Thus ordinary and composite channel-symmetry breakings are detected by 
even- and odd-frequency susceptibilities, respectively, both of which can be calculated from the two-particle Green function.

We can apply the same procedure to detect the composite pairing state.
Among various internal symmetries of the pairing, the most stable is
the channel-singlet/pseudospin-singlet pairing \cite{hoshino14}.
Hence, the relevant quantity is now
\begin{align}
p^+ (\tau, \tau') = \sum_{i\al\al'\sg\sg'} c^\dg_{i\al\sg}(\tau) \epsilon_{\al\al'} \epsilon_{\sg\sg'} c^\dg_{i\al'\sg'} (\tau') \, \epn^{\imu \bm Q\cdot \bm R_i}
\end{align}
instead of $m^z$.

We have already evaluated numerically both diagonal and off-diagonal odd-frequency susceptibilities by combining the continuous-time quantum Monte Carlo (CT-QMC) and the dynamical mean-field theory (DMFT)\cite{hoshino11, hoshino14}.
We have indeed found the divergence corresponding to the onset of diagonal and off-diagonal composite orders.

\section{Effective Hamiltonian with pseudofermions}
From numerical results \cite{hoshino11, hoshino14},
we have realized that the low-energy states in the ordered phase can be described in terms of fermionic excitations that originate not only from conduction bands but also from localized states forming the non-Kramers doublet.
Analogous situation has long been recognized in the standard Kondo lattice, where the local spins acquire the fermionic character by the Kondo effect.
The paramagnetic fixed point of the Kondo lattice can be 
described 
by 
hybridized bands with strongly renormalized parameters, 
which comes from Kondo resonance at each site.

Let us first construct the effective Hamiltonian for the diagonal composite order described by $\Psi^z$ of Eq.~\eqref{Psi^z}.  In the ordered phase, the conduction electrons in channel $\al=1$ forms a Kondo insulator with hybridized band, while the channel $\al=2$ remains a Fermi liquid decoupled form localized pseudospins.
This difference between the two channels is reflected in the order parameter $\la \Psi^z\ra = 2\sum_{i} \la \bm S_i \cdot (\bm s_{ci1} - \bm s_{ci2}) \ra$.
In analogy with the ordinary Kondo insulator \cite{riseborough, zhang},
we first rewrite the localized pseudospin operator as
\begin{align}
&\bm S_i = \frac 1 2 \sum_{i\sg\sg'} f^\dg_{i\sg} \bm \sg_{\sg\sg'} f_{i\sg'}
, \\
& \sum_{\sg} f^\dg_{i\sg} f_{i\sg} = 1
. \label{eq:local_constraint}
\end{align}
Note that fermionic operator $f_{i\sg}$ does not come from original $f$ electrons forming the non-Kramers doublet.  Hence we call these pseudofermions.
In the mean-field theory, the 
local operator constraint \eqref{eq:local_constraint} is replaced by a much looser constraint for the average of the operator.
In the composite order with $\Psi^z$, we replace the interaction term with channel $\alpha=1$ as
\begin{align}
J \bm S_i \cdot \bm s_{ci1} \longrightarrow 
V_1 \sum_\sg f^\dg_{i\sg} c_{i1\sg} + {\rm h.c. }
\end{align}
where the mean field
\begin{align}
V_1= - J \la c^\dg_{i1\sg} f_{i\sg} \ra
\label{V_1}
\end{align}
is independent of the site.
On the other hand, there is no mean field from the interaction
$J \bm S_i \cdot \bm s_{ci2}$ because the channel 2 does not form the hybridized band.
Note that 
we would obtain $3J/4$ instead of $J$ in Eq.~\eqref{V_1} in the faithful decoupling procedure.  
However we prefer the present choice since it gives the correct Kondo scale \cite{lacroix-cyrot1983}.

In this way we obtain the mean-field Hamiltonian
\begin{align}
{\cal H}_{\Psi^z} &= 
\sum_{\bm k\al\sg} \ep_{\bm k}c^\dg_{\bm k\al\sg} c_{\bm k\al\sg}
+ V_1\sum_{\bm k\al\sg } \left( f^\dg_{\bm k\sg} c_{\bm k1\sg} + c^\dg_{\bm k1\sg} f_{\bm k\sg} \right)
, \label{eq:MF_ham}
\end{align}
where the operators in the momentum space are defined by the Fourier transform of the real-space operators.
The single-particle spectra consist of three branches, which are given by
\begin{align}
E_{\bm k\pm} &= \frac 1 2 \left[ \ep_{\bm k} \pm \sqrt{\ep_{\bm k}^2 + 4V_1^2} \right]
, \label{eq:sp_energy1}
\\
E_{\bm k0} &= \ep_{\bm k}
, \label{eq:sp_energy2}
\end{align}
where $E_{\bm k0}$ describes quasi-particles in channel $\al=2$.
These branches qualitatively reproduce the results obtained by the DMFT at low energies.
If we solve the mean-field equation \eqref{V_1} self-consistently, the characteristic energy scale $\Gamma$ such as the transition temperature and ground-state energy is obtained as $\rho (0) \Gamma \sim \epn^{- 1/\rho(0) J}$ where $\rho(0)$ is the density of states at the Fermi level.
This gives usual expression for the Kondo temperature.
We note that the approximation here is valid only at sufficiently low temperatures, since the disordered state of the two-channel Kondo lattice is a non-Fermi liquid which cannot be described by the present theory.

Next we turn to the effective Hamiltonian in the ordered phase with $\Phi^x$.
Because of the SO(5) symmetry, which applies not only to the ground state but all excited states, the total single-particle spectrum 
\begin{align}
\rho (\bm k,\ep) = -\frac 1{2\pi} 
\imag\, {\rm Tr}\, \hat{\bf G}(\bm k, \ep+\imu\eta)
\end{align}
is the same in ordered states either with $\Psi^z$ or $\Phi^x$. Here the Green function 
$\hat{\bf G}(\bm k, \ep+\imu\eta)$ is an $8\times8$ matrix in the generalized Nambu space with components; 
$(c^\dagger _{\al\sg}, c_{\al'\sg'})$ with $\al,\al' =1,2$ and $\sg,\sg' =\pm 1$.

To be more specific, using the unitary transformations defined in Eqs.~(\ref{eq:unitary_transf1}--\ref{eq:unitary_transf3}), we can derive the effective Hamiltonian for composite pairing state.
Let us first regard $V_1$ as independent of $\mathscr{U}$.
Then the result of transformation is given by
\begin{align}
\mathscr{U} {\cal H}_{\Psi^z} \mathscr{U}^{-1} 
\equiv {\cal H}_{\Phi^x} 
= \sum_{\bm k\al\sg} \ep_{\bm k}c^\dg_{\bm k\al\sg} c_{\bm k\al\sg}
+ \frac{V_1}{\sqrt 2}\sum_{\bm k\sg\sg' } \left(
  \delta_{\sg\sg'} f^\dg_{\bm k\sg} c_{\bm k1\sg}
+ \epsilon_{\sg\sg'} f^\dg_{\bm k\sg} c^\dg_{-\bm k-\bm Q, 2\sg'}
 + {\rm h.c.} \right).
\end{align}
If we start from the original model given by Eq.~\eqref{2chKL}, 
we may decouple the interaction part so as to allow for 
a new nonzero average such as $\la c_{i 2\sg} f_{i\sg} \ra \neq 0$.  
By setting equal magnitudes to diagonal and off-diagonal 
mean fields, we obtain the result identical to
${\cal H}_{\Phi^x}$. 
As can be checked by direct calculation, the single-particle spectra of ${\cal H}_{\Phi^x}$
are the same as Eqs.~\eqref{eq:sp_energy1} and \eqref{eq:sp_energy2}.
However, the wave function of each excitation is of course different.  As a result, 
${\cal H}_{\Psi^z} $ and ${\cal H}_{\Phi^x}$ have the same specific heat, but show
very different electromagnetic properties.

\section{Collective excitation modes}
Now we discuss collective excitations in the composite channel ordered state with $\Psi^z$.
Within the equation of motion method, we include self-consistently induced field  that corresponds to vertex corrections in  the Bethe-Salpeter equation \cite{anderson}.
In order to derive collective excitations 
from the ordered state,
it is convenient to add an (infinitesimal) channel flipping field $h_c$ with wave number $\bm q$:
\begin{align}
{\cal H}_{\rm ext}(\bm q) &= - h_c \sum_{\bm k\sg } \left( c^\dg_{\bm k+\bm q,1 \sg} c_{\bm k2\sg} + c^\dg_{\bm k2\sg} c_{\bm k+\bm q,1 \sg} \right) .
\end{align}
The channel field in the non-Kramers doublet corresponds to a modulating magnetic field 
in the $xy$-plane.
With this external field,  hybridization in channel $\al=2$ is induced infinitesimally.
In the random phase approximation, 
the induced hybridization is accounted for by the Hamiltonian 
\begin{align}
{\cal H}_{\rm ind}(\bm q) &= V_2 \sum_{\bm k\sg } \left( f^\dg_{\bm k+\bm q, \sg} c_{\bm k2\sg} + c^\dg_{\bm k2\sg} f_{\bm k+\bm q, \sg} \right), \\
V_2 = &-\frac{J}{N} \sum_{\bm k} \la c^\dg_{\bm k2\sg} f_{\bm k+\bm q,\sg} \ra
,
\end{align}
where $N = \sum_i 1$ is the total number of sites, and $V_2 \sim O(h_c)$.
With ${\cal H} = {\cal H}_{\Psi^z} + {\cal H}_{\rm ext} + {\cal H}_{\rm ind}$,
the Heisenberg equations of motion $\imu\partial_t {\cal O} = [{\cal O}, {\cal H}]$ 
are explicitly written as
\begin{align}
\imu \partial_t (c^\dg_{\bm k+\bm q, 1\sg} c_{\bm k2\sg})
 &= 
(- \ep_{\bm k + \bm q}+\ep_{\bm k}) c^\dg_{\bm k+\bm q, 1\sg} c_{\bm k2\sg}
- V_1 f^\dg_{\bm k+\bm q, \sg} c_{\bm k2\sg}
+ V_2 c^\dg_{\bm k+\bm q, 1\sg} f_{\bm k+\bm q, \sg}
\nonumber \\
&\ \ + h_c c^\dg_{\bm k 2\sg} c_{\bm k2\sg}
- h_c c^\dg_{\bm k+\bm q, 1\sg} c_{\bm k+\bm q, 1\sg}
, \\
\imu \partial_t (f^\dg_{\bm k+\bm q, \sg} c_{\bm k2\sg}) &= 
\ep_{\bm k} f^\dg_{\bm k+\bm q, \sg} c_{\bm k2\sg}
- V_1 c^\dg_{\bm k+\bm q, 1\sg} c_{\bm k2\sg}
- V_2 c^\dg_{\bm k 2\sg} c_{\bm k2\sg}
+ V_2 f^\dg_{\bm k+\bm q, \sg} f_{\bm k+\bm q, \sg}
\nonumber \\
&\ \ - h_c f^\dg_{\bm k+\bm q, \sg} c_{\bm k+\bm q, 1\sg}
. \end{align}
If the external field is oscillating with frequency $\omega$, the Fourier components are determined by
\begin{align}
&\begin{pmatrix}
\la c^\dg_{1\sg} c_{2\sg} \ra (\bm q, \omega) \\
\la f^\dg_{\sg} c_{2\sg} \ra (\bm q, \omega)
\end{pmatrix}
= 
\begin{pmatrix}
\chi^{11}_{\bm q}(\omega) & \chi^{12}_{\bm q}(\omega)\\
\chi^{21}_{\bm q}(\omega) & \chi^{22}_{\bm q}(\omega)\\
\end{pmatrix}
\begin{pmatrix}
h_c \\
-V_2
\end{pmatrix}, 
\\
&\begin{pmatrix}
\chi^{11} & \chi^{12}\\
\chi^{21} & \chi^{22}\\
\end{pmatrix}
= - \frac 1 N \sum_{\bm k}
\begin{pmatrix}
\omega+\ep_{\bm k+\bm q}-\ep_{\bm k} & V_1\\
V_1 & \omega - \ep_{\bm k} \\
\end{pmatrix}^{-1}
\begin{pmatrix}
 n_{c1, \bm k+\bm q} - n_{c2\bm k} & X_{\bm k+\bm q}\\
X_{\bm k+\bm q} & n_{f, \bm k+\bm q} - n_{c2\bm k} 
\end{pmatrix},
\label{eq:def_chi}
\end{align}
where the quantities in the RHS are evaluated from the Hamiltonian ${\cal H}_{\Psi^z}$ as
\begin{align}
X_{\bm k} &= \la f^\dg_{\bm k\sg} c_{\bm k1\sg} \ra
= V_1 [f(E_{\bm k+}) - f(E_{\bm k-})] / (E_{\bm k+}-E_{\bm k-})
, \\
n_{f\bm k} &= \la f^\dg_{\bm k\sg} f_{\bm k\sg} \ra
= [- E_{\bm k-}f(E_{\bm k+}) + E_{\bm k+} f(E_{\bm k-})] / (E_{\bm k+}-E_{\bm k-})
, \\
n_{c1\bm k} &= \la c^\dg_{\bm k1\sg} c_{\bm k1\sg} \ra
= [E_{\bm k+}f(E_{\bm k+}) - E_{\bm k-}f(E_{\bm k-})] / (E_{\bm k+}-E_{\bm k-})
, \\
n_{c2\bm k} &= \la c^\dg_{\bm k2\sg} c_{\bm k2\sg} \ra
= f(E_{\bm k0})
,
\end{align}
with the Fermi distribution function $f(\ep) = 1 / (\epn^{\beta\ep}+1)$.

Using these quantities, the relevant dynamical susceptibility is derived as 
\begin{align}
\chi_{\bm q}(\omega) &= \la c^\dg_{1\sg} c_{2\sg} \ra (\bm q, \omega)/h_c = 
\frac{\chi^{11} - J (\chi^{11}\chi^{22} - \chi^{12}\chi^{21})}
{1 - J \chi^{22}}
, \label{eq:dynamical_sus1} 
\end{align}
whose pole determines the dispersion relation of the Goldstone modes.
Assuming small $|{\bm q}|$ and $\omega$, we expand the denominator as
\begin{align}
1-J\chi^{22}_{\bm q} (\omega) 
&= c_0 (\bm q) 
+ c_2 (\bm q) \omega^2 + O(\omega^4)
. \label{eq:expand_pole}
\end{align}
The odd-order terms of $\bm q$ and $\omega$ vanish because of the spatial inversion and particle-hole symmetries, respectively.
In the lowest-order with respect to $\bm q$ and $\omega$, 
each coefficient is evaluated at $T=0$ as
\begin{align}
c_0(\bm q) &=
- \frac{J}{2NV_1^4} \sum_{\bm k} (\bm v_{\bm k} \cdot \bm q)^2\left[
\frac{\ep_{\bm k}^4 + 6V_1^2\ep_{\bm k}^2 + 6V_1^4}{(\ep_{\bm k}^2 + 4V_1^2)^{3/2}}
- |\ep_{\bm k}| - V_1^2 \delta (\ep_{\bm k})
\right]
+ O(\bm q^4), 
\label{eq:c0}
\\
c_2(\bm q) &=
\frac{J}{2NV_1^4} \sum_{\bm k} \left[
|\ep_{\bm k}| - \frac{\ep_{\bm k}^2 + 2V_1^2}{\sqrt{\ep_{\bm k}^2 + 4V_1^2}}
\right]
+O(\bm q^2).
\end{align}
Here we have defined the velocity by $\bm v_{\bm k} = \partial \ep_{\bm k} / \partial \bm k$.
For sufficiently small $V_1$, the integrands are sharply peaked at the Fermi level, and hence we can represent the density of states by its values near the Fermi level.
Namely, we use the following approximations
\begin{align}
\tilde \rho (\ep) &= \frac 1 N \sum_{\bm k} \bm v_{\bm k}^2 \delta (\ep - \ep_{\bm k}) = \tilde \rho (0) + \frac 1 2 \tilde \rho''(0) \ep^2 + O(\ep^4)
,\label{eq:v-DOS} \\
\rho (\ep) &= \frac 1 N \sum_{\bm k} \delta (\ep - \ep_{\bm k}) = \rho (0) + O(\ep^2)
.\label{eq:DOS}
\end{align}
In Eq.~\eqref{eq:v-DOS}, we need terms up to $O(\ep^2)$ since
contribution of $\tilde \rho (0)$ is cancelled in Eq.~\eqref{eq:c0}.
Thus the energy integration can be performed.
We note that $\tilde \rho''(0)$ is 
 estimated as $\tilde \rho''(0) \sim - v_{\rm F}^2 \rho(0) / D^2$ with $v_{\rm F}^2 = \la \bm v_{\bm k}^2 \ra_{\rm FS}$ being 
averaged on the Fermi surface and $D$ being the energy with the order of the band width.

The dispersion relation $\omega = \omega_{\bm q}$ is determined by putting Eq.~\eqref{eq:expand_pole} equal to zero, 
and we obtain
\begin{align}
\omega_{\bm q} = \frac{v_{\rm F}}{\sqrt{d}} \left( \frac{V_1}{2D} \right) \, |\bm q|
, \label{eq:gs_dispersion}
\end{align}
where $d=3$ is the dimension of the system.
The velocity of the collective modes is proportional to the Fermi velocity and the magnitude of the mean-field.
The factor $V_1/D$ arises due to hybridization between the conduction electrons and localized pseudofermions, which modifies the Fermi velocity $v_{\rm F}$ of conduction electrons.

The combination of the Fermi velocity and the order parameter in giving the velocity is a characteristic feature in the present ordered state.
Namely, in itinerant magnetism or superfluid in the weak-coupling limit, the velocity of the Goldstone mode is close to the Fermi velocity. 
On the other hand, the velocity in localized antiferromagnet is proportional to the exchange interaction times the spontaneous magnetization within the mean-field theory.

\begin{figure}[t]
\begin{center}
\includegraphics[width=60mm]{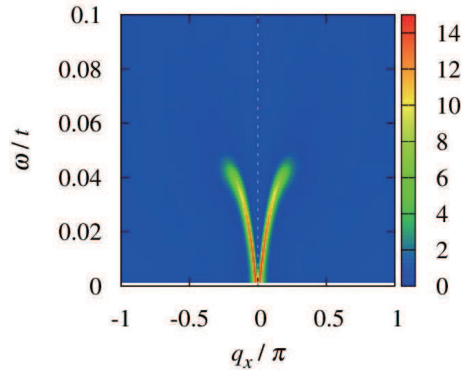}
\caption{(Color online)
Two-particle spectrum probed by $\imag \chi_{\bm q = (q_x, 0, 0)} (\omega+\imu\eta)/\omega$.
The parameters are chosen as $V_1 = 0.5$, $T=0.025$ and $\eta = 0.00375$ with $t=1$ being the unit of energy.
}
\label{fig:gs_mode}
\end{center}
\end{figure}

The two-particle spectrum appears as peaked features 
in $\imag \chi_{\bm q}(\omega+\imu\eta)/\omega$ if scanned over the $(\bm q,\omega)$ plane.
We assume the simple cubic lattice in three dimension, where the conduction band is described by
\begin{align}
\ep_{\bm k} = -2t (\cos k_x + \cos k_y + \cos k_z)
\end{align}
with the lattice constant being unity.
Figure \ref{fig:gs_mode} shows the two-particle spectra calculated from Eq.~\eqref{eq:dynamical_sus1}.
We can clearly see the low-lying collective mode located around $\bm q=\bm 0$.
Experimentally, the channel moment corresponds to the spin magnetic moment of conduction electrons.
Hence, if inelastic neutron scattering experiment is performed in a
system with the composite channel order,
the magnetic spectrum should look like Fig.~\ref{fig:gs_mode}.

On the other hand, for composite pairing with the staggered ordering vector $\bm Q$, the spectrum of the Goldstone mode is derived simply by shifting the momentum as $\bm q \rightarrow \bm q + \bm Q$,  provided the system is at half filling.
The resultant dispersion relation $\omega = \omega'_{\bm q}$ is given by
\begin{align}
\omega'_{\bm q} = \omega_{\bm q - \bm Q}
\end{align}
where the RHS is given by Eq.~\eqref{eq:gs_dispersion} for small $|\bm q-\bm Q|$.
Namely, the intensity map in the $(\bm q,\omega)$ space should 
again 
look like Fig.~\ref{fig:gs_mode} but the branch arises at $\bm q = \bm Q$.
The collective mode in the composite pairing state can be detectable in the charge part of the dynamical susceptibilities.

It is well known for ordinary superconductors that the charge Goldstone mode is absorbed into a plasmon by the long-ranged Coulomb interaction \cite{anderson}, which has singularity in the limit $\bm q\rightarrow \bm 0$.
In the present staggered pairing state, on the other hand, 
the Coulomb interaction has no singularity at $\bm q = \bm Q$.
Hence it is expected that the coupling of the collective excitation with the plasmon mode is weaker, or negligible.
At low temperature, 
the inelastic light scattering experiment is one of candidates to detect this type of off-diagonal order.

\section{Summary and Discussion}

We have demonstrated in this paper that onset of the composite order parameter 
can be detected from divergence of the odd-frequency susceptibility, which is derived within the standard framework of two-particle Green functions.
We have also shown that a mean-field theory with pseudofermions provides a much simpler description of composite orders in the two-channel Kondo lattice. 
The simplicity of the framework has made it possible to derive the spectrum of the collective excitations from the ordered state.

In actual non-Kramers doublet systems, the particle-hole symmetry in the conduction bands can hardly be realized.
In this case the diagonal composite order accompanies 
ferromagnetic moment of conduction electrons \cite{hoshino11}.
The magnitude of the moment depends on degree of the broken particle-hole symmetry,
but is expected to be tiny in general.
Note that the present collective mode in the presence of spontaneous moment is different from the ordinary ferromagnetic magnon mode.
Namely, 
the spectrum 
$\omega_{\bm q} \propto |\bm q|$ is in contrast with  
the magnon spectrum $\omega_{\bm q} \propto \bm q^2$. 
The latter is common to localized and itinerant ferromagnets with spin SU(2) symmetry.
If there is a magnetic anisotropy caused, for example,  by spin-orbit interaction in conduction bands, the collective mode in the composite diagonal order should acquire a gap in the spectrum.

On the other hand, it has been demonstrated \cite{hoshino13,hoshino14} that the composite superconductivity is more stable than the diagonal order in the density range without the particle-hole symmetry.  Hence we expect that actual competition in real materials should be between the ordinary quadrupole order by the RKKY interaction, and the composite superconductivity.
If the composite superconductivity is indeed realized in non-Kramers doublet systems, the Goldstone mode should remain gapless even without the particle-hole symmetry.
Furthermore, the gapless point corresponds to the boundary of the Brillouin zone. 
In the smaller Brillouin zone associated with the staggered order,  the zone boundary is folded to the center of the Brillouin zone.
It remains to see how the zone-folding effect affects the physical property of the collective mode.
We expect that this collective mode may be probed by inelastic light scattering and other measurement that detects excitations with charge degrees of freedom.
The observation clearly identifies the composite superconductivity, as distinct from the ordinary one.

\end{document}